\documentclass[aps,prb,twocolumn,showpacs,amsfonts,superscriptaddress]{revtex4-1}
\usepackage{epsfig}\usepackage{subfigure}
\usepackage{graphicx,graphics}
\usepackage{amsmath}
\usepackage{amssymb}
\usepackage{bm}
\usepackage{color}
\usepackage[utf8]{inputenc}

\begin{document}

\title{Andreev Spectroscopy of Nonhelical Spin Textures in Topological Insulators}

\author{David J. Alspaugh}
\email[]{Current address: Department of Physics and Astronomy, California State University, Northridge, CA 91330; david.alspaugh@csun.edu}
\affiliation{Department of Physics and Astronomy, Louisiana State University, Baton Rouge, LA 70803-4001}

\author{Mahmoud M. Asmar}
\email[]{masmar@kennesaw.edu}
\affiliation{Department of Physics and Astronomy, Louisiana State University, Baton Rouge, LA 70803-4001}
\affiliation{Department of Physics, Kennesaw State University, Marietta, GA 30060}

\author{Daniel E. Sheehy}
\email[]{sheehy@lsu.edu}
\affiliation{Department of Physics and Astronomy, Louisiana State University, Baton Rouge, LA 70803-4001}

\author{Ilya Vekhter}
\email[]{vekhter@lsu.edu}
\affiliation{Department of Physics and Astronomy, Louisiana State University, Baton Rouge, LA 70803-4001}

\date{\today}

\begin{abstract}

{We study how nonhelical spin textures affect the proximity-induced superconductivity of topological insulator (TI)-superconductor (SC) interface states. In particular we calculate the conductance of lateral heterojunctions which comprise a TI surface that is only partially covered by a superconducting material. Interface potentials at the TI-SC interface may lead to a Fermi velocity and spin texture mismatch between the two regions of the lateral heterojunction. By enforcing the hermiticity of the total Hamiltonian, we derive the boundary conditions and calculate the conductance of the structure in both the normal and superconducting state. The total Andreev conductance is calculated for both $s$-wave and spin-triplet parent SCs, and for several examples of nonhelical spin textures which lead to different Fermi surface mismatches between the two planar regions of the heterojunction. We find that for spin-triplet SCs, nonzero conductance signatures only appear for certain combinations of nonhelical spin textures and parent superconducting material.}

\end{abstract}


\maketitle

\section{Introduction}
Topological Insulators (TIs) are a class of materials which admit linearly dispersing surface states and preserve time reversal (TR) symmetry.~\cite{Hasan2010,Qi2011} At pristine vacuum terminations these surface states have isotropic dispersions and are perfectly helical, such that the spin of the propagating state is perpendicular to the direction of its momentum and is confined within the plane of the termination. The common assumption is that at interfaces of TIs with other materials, the interface states of the TI exhibit these same properties. However, the properties of the interfacial boundary may be qualitatively different than that of a vacuum terminated surface, and the available set of symmetries which the interface states must obey is lower than that of the bulk TI material. Effects from charge redistribution, dangling bonds, and lattice mismatch can introduce TR-preserving interface potentials which lower the symmetry and consequently alter the spin structure and dispersion of the interface states. These states have been shown to exhibit elliptical energy contours and nonhelical spin-momentum locking.~\cite{Asmar2017,Alspaugh2018,Zhang2012,Thareja2020}

\begin{figure}
\centering
\includegraphics[scale=0.8]{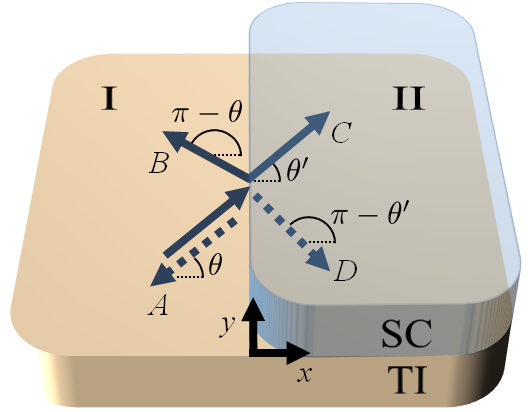}
\caption{Schematic of the superconducting lateral heterojunction. The device is created by placing only half of a TI surface in contact with a superconducting material. Region I on the $x<0$ half-space defines the TI-vacuum surface, and region II on the $x>0$ half-space defines the TI-SC interface. Arrows indicate the possible scattering events of normal and Andreev reflection, which determine the conductance of the device. For an incoming electron with angle $\theta$ towards the $x = 0$ boundary, these possibilities include Andreev reflection as a hole ($A$), specular reflection ($B$), transmission as an electron-like quasiparticle ($C$), and transmission as a hole-like quasiparticle ($D$).}
\label{schematic}
\end{figure}

Some of the most exciting potential applications of TIs focus on the possibility that they may be useful as platforms for performing fault-tolerant quantum computations. For instance, If these materials are placed in contact with superconductors (SCs), the interface state of the TI experiences superconductivity due to the proximity effect. When vortices are present or when placed alongside ferromagnetic systems, these junctions are predicted to host Majorana zero modes.~\cite{Fu2008} These objects are non-abelians anyons which have the ability to encode quantum information as they are braided around one-another in space.~\cite{Fu2008,Read2000,Ivanov2001} However, the realization of Majorana zero modes in experiment has proved elusive.~\cite{Zhang2018,Zhang2021} These difficulties emphasize the need for a detailed understanding of the physics of topological interfaces. In the case of TI-SC heterostructures, the properties of the induced superconductivity strongly depend on the spin structure of the interface state as well as the properties of the parent SC. Existing conclusions about the interface superconductivity, including the prediction that some parent spin-triplet SCs do not induce superconductivity at the interface at all, have been reached assuming helical TI surface states. In Ref.~\onlinecite{Alspaugh2018}, it was demonstrated that nonhelical spin textures are required in order to observe proximity-induced superconductivity for several classes of parent SCs.

To study the consequences of junctions on the TI interface state, in this work we analyze the conductance signatures of lateral heterojunctions of TI-vacuum and TI-SC interfaces. As shown schematically in Fig.~\ref{schematic}, such lateral heterojunctions can be constructed by covering only part of a TI surface with a parent SC, where here we denote the TI-vacuum interface on the $x < 0$ half space as region I and the TI-SC interface on the $x > 0$ half space as region II. While the TI-vacuum interface has perfectly helical surface states, interface potentials between the TI and SC materials result in nonhelical interface states. These new states with their altered spin textures have different anisotropic Fermi velocities compared to the helical states, making the analysis of their junctions nontrivial. In Sec.~\ref{sec2} we introduce the model for the lateral heterojunction, and by enforcing the hermiticity of the Hamiltonian across both regions I and II in Fig.~\ref{schematic}, we develop a formalism allowing us to calculate the normal state conductance across the device. In Sec.~\ref{sec3} we introduce superconductivity by placing an $s$-wave SC in region II and calculate the superconducting Andreev conductance across the lateral heterojunction. In Sec.~\ref{sec4} we analyze these results for several different experimental setups. We compare how the scattering across the heterojunction described by our formalism compares to the effects of a mismatch in the chemical potential between regions I and II, and we analyze the Andreev conductance arising from several different nonhelical spin textures within region II. In Sec.~\ref{sec5} we modify our model from Sec.~\ref{sec3} by introducing spin-triplet SCs within region II, and we analyze the Andreev conductance for parent spin-triplet SCs from the $D_{4\rm h}$ point group. Critically, we find that for certain types of spin-triplet SCs, nonzero conductance signatures exist only if nonhelical spin textures are present in region II. In Sec.~\ref{sec6} we summarize our main results and discuss the key experimental consequences of nonhelical spin textures.

\section{Model of the Lateral Heterojunction}
\label{sec2}
We begin with a description of the lateral heterojunction in the normal state. As shown in Fig.~\ref{schematic}, this device can be created by covering half of a TI surface with a superconducting material. Here we define the $x < 0$ half-space with the TI-vacuum surface as region I, and the $x > 0$ half-space with the TI-SC interface as region II. While region I will host a helical TI surface state with surface state spins confined to the surface (i.e. in the $x$-$y$ plane of Fig.~\ref{schematic}), the material junction in region II can result in interface states with nonhelical spin textures, giving rise to elliptical energy dispersions and spins with components that point out of the plane of the interface (i.e., in the $z$-direction of Fig.~\ref{schematic}).~\cite{Asmar2017,Alspaugh2018} To study the heterojunction we therefore write a helical surface Hamiltonian in region I and include the most general effective linear Hamiltonian that is TR invariant in region II,
\begin{equation}
H = \begin{cases}
\hbar v_{\rm F}(\bm{\sigma}\times -i\bm{\nabla})_{z} - \mu_{L}, & \text{for} \ x<0
\\ \bm{c}\cdot \bm{\sigma} - \mu_{R}, & \text{for} \ x > 0
\end{cases}.
\label{latmodel}
\end{equation}
Here $\bm{\sigma} = (\sigma_{x},\sigma_{y},\sigma_{z})^{T}$ is a vector of Pauli matrices in spin-space, $v_{\rm F}$ is the Fermi velocity, $\mu_{L}$ and $\mu_{R}$ are the chemical potentials in regions I and II respectively, and $\bm{c}$ is a three-component vector defined by $c_{i} = -i\sum_{j=x,y}c_{ij}\partial_{j}$. We notice that in the case $\bm{c}_{\rm D} = -i\hbar v_{F}(\partial_{y},-\partial_{x},0)^{T}$, the Hamiltonians of regions I and II equal each other and we recover a single TI surface over all space. The presence of nonhelical spin textures in the TI-SC interface of region II are can then be encoded in the choice of $c_{ij}$ coefficients.~\cite{Asmar2017} This prescription is valid so long as the transmission between the TI and SC materials is low, so that the TI interface state is still well-defined.~\cite{Zhao2010,Grein2012} We also note that from the construction of the vector $\bm{c}$, the form of the Hamiltonian in region II is identical to that of antisymmetric spin-orbit coupling in noncentrosymmetric metals, whose consequences on superconductivity have been extensively studied.~\cite{Smidman2017,Samokhin2015} What is different however is that the usual quadratic energy term is absent, placing us in a regime of infinite spin-orbit coupling strength.~\cite{Pesin2012}

\subsection{Normal State Conductance}
In this section we calculate the normal state conductance of the lateral heterojunction described by Eq.~\eqref{latmodel}. To do this, we must solve the scattering problem in which incoming electrons in region I of Fig.~\ref{schematic} may either specularly reflect at the $x = 0$ boundary or transmit into region II. Matching the wave functions at the boundary is nontrivial, however, as the Hamiltonians of the two regions have different Fermi velocities. Our goal then is to match the wave functions in a way that is consistent with the hermiticity of the total Hamiltonian in Eq.~\eqref{latmodel}.

The effective long-wavelength description cannot account for the rapid variations of the wave functions in the vicinity of the potential edge at $x = 0$. Therefore, due to the fact that we only have a linear description, in this approach the envelope wave function is not continuous.~\cite{Ahari2016,Bonneau2001,Gitman2012,Stone2009} Instead it satisfies
\begin{equation}
\psi_{\rm I}(0,y) = \mathcal{M}\psi_{\rm II}(0,y),
\label{boundarymatching}
\end{equation} 
for some matrix $\mathcal{M}$ and wave functions $\psi_{\rm I}(\bm{r})$ and $\psi_{\rm II}(\bm{r})$ in regions I and II respectively. In order for this to be a valid boundary value relation for our system, and hence valid for all wave functions of the Hilbert space, $\mathcal{M}$ must respect the hermiticity of the Hamiltonian $H$ expressed in Eq.~\eqref{latmodel}. This ensures particle conservation and therefore continuity of the current normal to the boundary. That is, if both $\psi_{1}(\bm{r})$ and $\psi_{2}(\bm{r})$ are wave functions for all $x$ that belong to $H$, we must have the inner product relations $\langle\psi_{1}|H\psi_{2}\rangle = \langle H\psi_{1}|\psi_{2}\rangle$. Inserting Eq.~\eqref{boundarymatching} into this condition for hermiticity and integrating by parts, we find
\begin{equation}
-\hbar v_{\rm F}\mathcal{M}^{\dagger}\sigma_{y}\mathcal{M} = \sum_{i=x,y,z}c_{ix}\sigma_{x}.
\label{hermiticity}
\end{equation}
To explicitly solve for each element in $\mathcal{M}$, it is helpful to restrict the form of $\mathcal{M}$ by requiring that it must also respect the discrete symmetries of the system. We let $\psi(\bm{r})$ be a wave function over all space such that
\begin{equation}
\psi(\bm{r}) = \begin{cases} \psi_{\rm I}(\bm{r}) & \text{for} \ x<0 
\\ \psi_{\rm II}(\bm{r}) & \text{for} \ x>0
\end{cases},
\end{equation}
and write $\mathcal{T} = i\sigma_{y}K$ as the TR symmetry operator where $K$ is complex conjugation. Because the system preserves TR symmetry, we know that $\mathcal{T}\psi(\bm{r})$ is also a wave function of our system. Eq.~\eqref{boundarymatching} holds for all wave functions in the Hilbert space, which implies both $\psi_{\rm I}(0,y) = \mathcal{M}\psi_{\rm II}(0,y)$ and $\mathcal{T}\psi_{\rm I}(0,y) = \mathcal{M}\mathcal{T}\psi_{\rm II}(0,y)$. However, by applying $\mathcal{T}$ on the left of both sides of Eq.~\eqref{boundarymatching}, we have $\mathcal{T}\psi_{\rm I}(0,y) = \mathcal{T}\mathcal{M}\psi_{\rm II}(0,y)$. This demonstrates that the commutator $[\mathcal{T},\mathcal{M}] = 0$. This condition implies that $\mathcal{M}$ must take the form $\mathcal{M} = \gamma_{0}\sigma_{0} + i(\gamma_{x}\sigma_{x} + \gamma_{y}\sigma_{y} + \gamma_{z}\sigma_{z})$, where all the $\gamma_{i}$ are purely real. With this restriction, Eq.~\eqref{hermiticity} can be solved to explicitly find each component of $\mathcal{M}$,
\begin{equation}
\mathcal{M}(\beta) = \sqrt{\dfrac{v}{v_{\rm F}}}\bigg[e^{i\sigma_{y}\beta} + \dfrac{i}{2\hbar v}(c_{xx}\sigma_{z} - c_{zx}\sigma_{x})e^{-i\sigma_{y}\beta}\bigg].
\label{matrixm}
\end{equation}
Here $v = (\sqrt{\sum_{i}c_{ix}^{2}} - c_{yx})/2\hbar$, and $\beta$ is a free parameter such that $\beta \in [0,2\pi)$.

Now that we can match the wave functions at the $x = 0$ boundary between regions I and II, the next step is to solve the scattering problem to obtain the normal state conductance of the lateral heterojunction. We consider an incoming electron in region I of Fig.~\ref{schematic} with momentum $ \bm{k}_{1} = (k_{x},k_{y})^{T}$ and in-plane momentum angle $\theta = \tan^{-1}k_{y}/k_{x}$, a specularly reflected electron with momentum $\bm{k}_{2} = (-k_{x},k_{y})^{T}$, and a transmitted electron in region II with momentum $\bm{k}_{1}^{\prime} = (k_{x}^{\prime},k_{y})^{T}$. The outgoing angle $\theta^{\prime}$ may be solved for in terms of the incoming angle due to the conservation of normal state energy and the conservation of the $k_{y}$ momentum. The wave function in region I is given by
\begin{equation}
\psi_{\rm I}(\bm{r}) = \dfrac{e^{i\bm{k}_{1}\cdot \bm{r}}}{\sqrt{2}}\begin{pmatrix}
1 \\ -ie^{i\theta}
\end{pmatrix}
 + r \dfrac{e^{i\bm{k}_{2}\cdot \bm{r}}}{\sqrt{2}}\begin{pmatrix}
 1 \\ ie^{-i\theta}
 \end{pmatrix}.
\end{equation}
In region II the wave function is
\begin{equation}
\psi_{\rm II}(\bm{r}) = t e^{i\bm{k}_{1}^{\prime}\cdot \bm{r}}  \begin{pmatrix}
\cos(\vartheta_{\bm{c}(\bm{k}_{1}^{\prime})}/2)
\\ e^{i\varphi_{\bm{c}(\bm{k}_{1}^{\prime})}}\sin(\vartheta_{\bm{c}(\bm{k}_{1}^{\prime})}/2)
\end{pmatrix}.
\end{equation}
Here $\vartheta_{\bm{c}(\bm{k}_{1}^{\prime})}$, $\varphi_{\bm{c}(\bm{k}_{1}^{\prime})}$ are the polar and azimuthal angles of the vector $\bm{c}(\bm{k}_{1}^{\prime})$ respectively, where $\bm{c}(\bm{k})$ is defined by $c_{i}(\bm{k}) = \sum_{j=x,y}c_{ij}k_{j}$, and $r$ and $t$ are the coefficients for the reflected and transmitted parts of the wave function respectively. Matching the wave functions at the $x = 0$ boundary by using Eq.~\eqref{boundarymatching}, we solve for the reflection coefficient $r$ and define the normal state conductance as
\begin{equation}
\sigma_{N}(\theta,\beta) = 1 - |r(\theta,\beta)|^{2}.
\label{sigman}
\end{equation}

\begin{figure}
\includegraphics[width=\columnwidth]{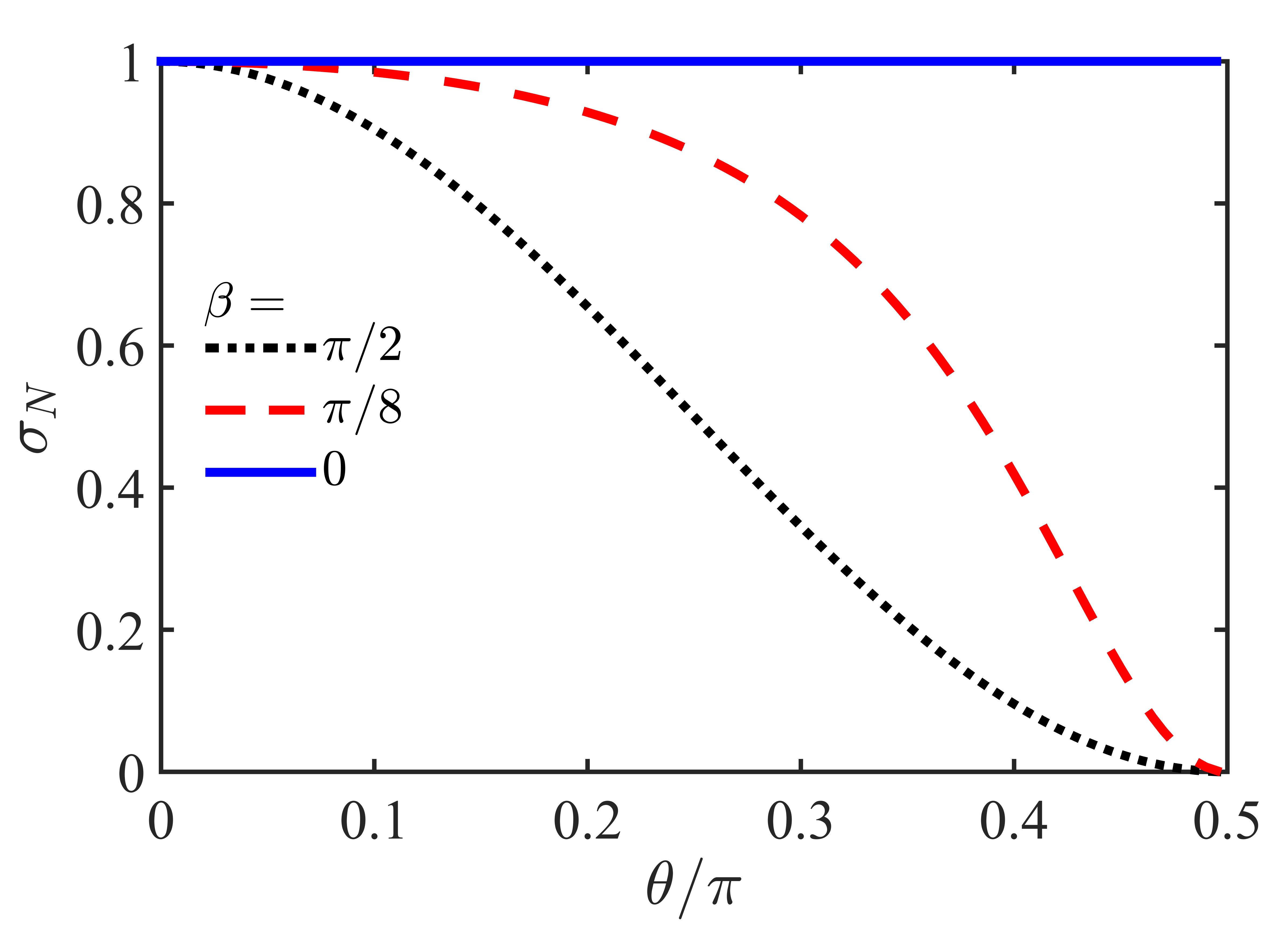}
\caption{Normal state conductance $\sigma_{N}(\theta,\beta)$ as given by Eq.~\eqref{sigman} for a lateral heterojunction between two helical surface Hamiltonians with the same Fermi velocity. We see that $\beta$ acts as an angle-dependent scattering amplitude at the one dimensional interface between regions I and II of the lateral heterojunction.}
\label{normalstateconductance}
\end{figure}

To demonstrate the consequences of the free parameter $\beta$ on the normal state conductance, we analyze $\sigma_{N}$ in the case when both regions I and II  are described by helical Hamiltonians and set $\bm{c} = \bm{c}_{\rm D}$ in Eq.~\eqref{latmodel}. When $\beta$ is zero, the conductance is always unity (in this case the heterojunction of Fig.~\ref{schematic} is equivalent to one single slab of a TI surface). However, when $\beta$ is nonzero the conductance decreases for nonzero angles. As shown in Fig.~\ref{normalstateconductance}, the most dramatic changes occur for $\beta = \pi/2$. We see that $\beta$ acts as an angle-dependent scattering amplitude localized at $x = 0$ between the two planar regions. Incoming electrons with $\theta = 0$ experience full transmission into region II as Dirac particles are incapable of back-scattering from normal incidence due to Klein tunneling.~\cite{Castro2009,Katsnelson2006,Allain2011,Asmar2014} This feature of Dirac particles is generally true for TR invariant systems as counter-propagating modes form Kramers' pairs. We can thus see that the effects of interfacial scattering potentials can be encoded into the form of the boundary matching matrix $\mathcal{M}(\beta)$.


\section{Superconducting Lateral Heterojunctions}
\label{sec3}
Having discussed the normal state of the lateral heterojunction, we now analyze the superconducting state. To calculate the conductance of the device shown in Fig.~\ref{schematic}, we first focus on region II and derive the proximity-induced superconducting order parameter of the TI-SC interface. We then use the Andreev equations to solve for the wave functions and find the conductance of the normal-superconducting junction of regions I and II.~\cite{Snelder2015,Blonder1982,Bruder1990,Kashiwaya1996} To obtain the conductance we solve the Andreev reflection scattering problem in which an incoming electron from region I may either be specularly reflected at the $x = 0$ boundary, or be retroreflected as a hole, corresponding to scattering events $B$ and $A$ in Fig.~\ref{schematic} respectively. Once again matching the wave functions is nontrivial, and we employ the boundary conditions derived above in the form suitable to treat superconductivity.

\begin{figure}
\includegraphics[width=\columnwidth]{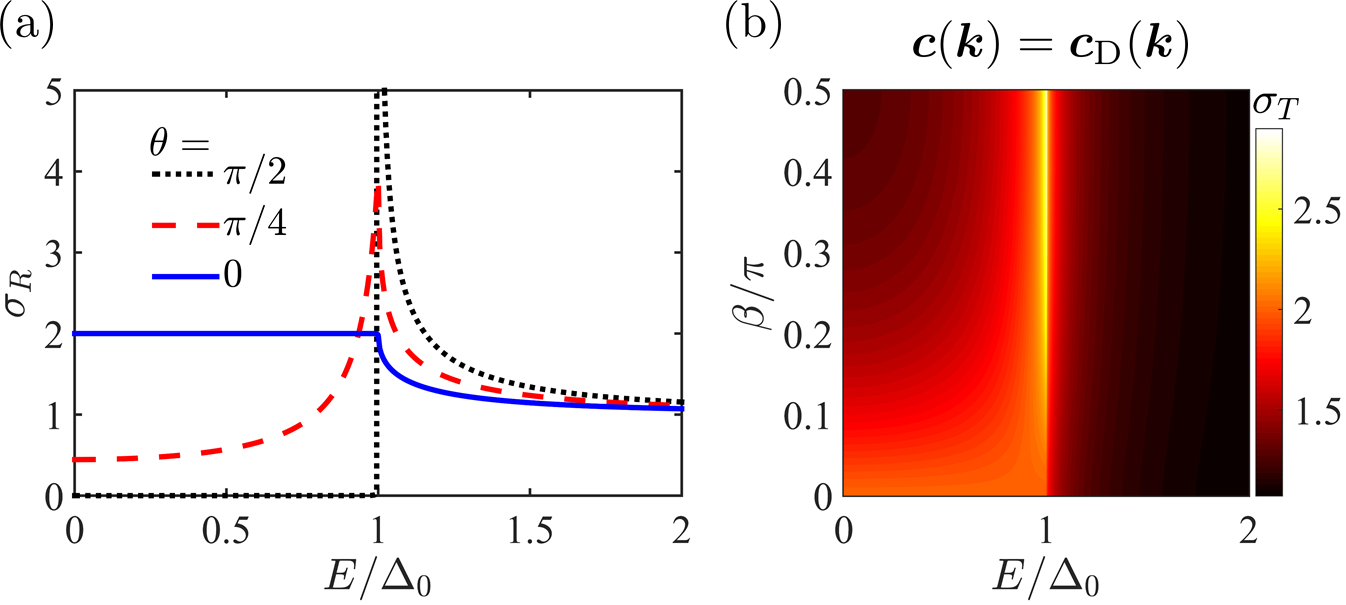}
\caption{Conductance of the superconducting lateral heterojunction for a parent $s$-wave SC. (a) Normalized conductance $\sigma_{R}(E,\theta,\beta)$ for several incident electron angles, such that $\bm{c}(\bm{k}) = \bm{c}_{\rm D}(\bm{k})$ and $\beta = \pi/2$. (b) Total conductance $\sigma_{T}(E,\beta)$ of the heterojunction such that $\bm{c}(\bm{k}) = \bm{c}_{\rm D}(\bm{k})$. The normalized conductance $\sigma_{R}(E,\theta,\beta)$ and total conductance $\sigma_{T}(E,\beta)$ are both defined in Eq.~\eqref{normtotcond}.}
\label{normcond}
\end{figure}

Focusing on region II, we can obtain the form of the proximity-induced order parameter of the TI-SC interface by projecting the Cooper pair structure of the parent SC onto the eigenstates of the interface Hamiltonian. Writing $(b_{\bm{k}\uparrow},b_{\bm{k}\downarrow})$ as the electron annihilation operators and introducing the Nambu-spinor $\Psi = (b_{\bm{k}\uparrow},b_{\bm{k}\downarrow},b_{-\bm{k}\uparrow}^{\dagger},b_{-\bm{k}\downarrow}^{\dagger})^{T}$, the Bogoliubov-de Gennes (BdG) Hamiltonian of the interface may be written as
\begin{equation}
\mathcal{H} = \dfrac{1}{2}\sum_{\bm{k}}\Psi^{\dagger}\begin{pmatrix}
H(\bm{k}) & \hat{\Delta}
\\ \hat{\Delta}^{\dagger} & -H^{T}(-\bm{k})
\end{pmatrix}
\Psi.
\label{bdgpsi}
\end{equation}
Here $H(\bm{k}) = \bm{c}(\bm{k})\cdot\bm{\sigma} - \mu_{R}$ is the interface Hamiltonian in the normal state, and the proximity-induced order parameter is given by $\hat{\Delta} = [\psi(\bm{k}) + \bm{d}(\bm{k})\cdot\bm{\sigma}](i\sigma_{y})$. The spin-singlet and spin-triplet components of the proximity-induced order are described by $\psi(\bm{k})$ and $\bm{d}(\bm{k})$ respectively. In this section, we shall focus on the case that the parent material of the heterostructure is an $s$-wave SC by setting $\psi(\bm{k}) = \Delta_{0}$ and $\bm{d}(\bm{k}) = 0$.

We then take the weak coupling limit to obtain the BdG wave functions.~\cite{Alspaugh2018} In the weak coupling limit where $\mu_{R} \gg \Delta_{0}$ interband pairing can be ignored. Writing $a_{\bm{k}}$ as the annihilation operator of the conduction band eigenstate of $H(\bm{k})$, we find
\begin{equation}
\mathcal{H} \approx \dfrac{1}{2}\sum_{\bm{k}}\Phi^{\dagger}\begin{pmatrix}
\xi(\bm{k}) & \Delta(\bm{k})
\\ \Delta^{*}(\bm{k}) & -\xi(\bm{k})
\end{pmatrix}
\Phi.
\label{bdgphi}
\end{equation}
Here $\Phi = (a_{\bm{k}},a_{-\bm{k}}^{\dagger})^{T}$ and $\Delta(\bm{k}) = -e^{-i\varphi_{\bm{c}(\bm{k})}}\Delta_{0}$, with $\varphi_{\bm{c}(\bm{k})}$ being the azimuthal angle of the vector $\bm{c}(\bm{k})$. The normal state energy is given as $\xi(\bm{k}) = |\bm{c}(\bm{k})| - \mu_{R}$. In the basis of Eq.~\eqref{bdgphi}, the eigenfunction of $\mathcal{H}$ has the form $\phi(\bm{r}) = (u(\bm{r}),v(\bm{r}))^{T}$. 

We may then invoke the Andreev approximation to model the boundary between regions I and II. For a spatially inhomogeneous order parameter, the BdG equations are given by~\cite{Bruder1990}
\begin{equation}
\begin{aligned}
E u(\bm{r}_{1}) &= \xi u(\bm{r}_{1}) + \int d\bm{r}_{2}\Delta(\bm{r}_{1},\bm{r}_{2})v(\bm{r}_{2}),
\\ E v(\bm{r}_{1}) &= -\xi v(\bm{r}_{1}) + \int d\bm{r}_{2}\Delta^{*}(\bm{r}_{1},\bm{r}_{2})u(\bm{r}_{2}).
\end{aligned}
\end{equation}
Here $\Delta(\bm{r}_{1},\bm{r}_{2})$ is the two-point correlation function. In the Andreev approximation, we write $\phi(\bm{r}) = e^{i\bm{k}_{\rm F}\cdot \bm{r}}[U(\bm{r}),V(\bm{r})]^{T}$, where $\bm{k}_{\rm F}$ lies on the Fermi surface. By writing $\bm{R} = (\bm{r}_{1} + \bm{r}_{2})/2$ and $\bm{s} = \bm{r}_{2} - \bm{r}_{1}$ as the center of mass and relative coordinates respectively, defining $\Delta(\bm{R},\bm{k}) = \int\Delta(\bm{R},\bm{s})e^{i\bm{k}\cdot \bm{s}}d\bm{s}$, and writing $\xi(\bm{k}_{\rm F} - i\bm{\nabla}) \approx (\partial \xi/\partial \bm{k})|_{\bm{k}_{\rm F}}\cdot(-i\bm{\nabla})$, we obtain the Andreev equations~\cite{Bruder1990}
\begin{equation}
\begin{aligned}
E U(\bm{r}_{1}) = -i\bm{v}_{\rm F}(\bm{k}_{\rm F})\cdot \bm{\nabla} U(\bm{r}_{1}) + \Delta(\bm{r}_{1},\bm{k}_{\rm F})V(\bm{r}_{1}),
\\ E V(\bm{r}_{1}) = i \bm{v}_{\rm F}(\bm{k}_{\rm F})\cdot \bm{\nabla} V(\bm{r}_{1}) + \Delta^{*}(\bm{r}_{1},\bm{k}_{\rm F})U(\bm{r}_{1}).
\end{aligned}
\end{equation}
Here $\bm{v}_{\rm F}(\bm{k}_{\rm F}) = (\partial\xi/\partial\bm{k})|_{\bm{k}_{\rm F}}$, and $\Delta(\bm{r}_{1},\bm{k}_{\rm F}) = -e^{-i\varphi_{\bm{c}(\bm{k}_{\rm F})}}\Delta_{0}\Theta(x)$. By solving these equations we can obtain the wave functions for our system. 

To solve the Andreev scattering problem, we assume a helical surface Hamiltonian in region I and consider an incoming incident electron with momentum $\bm{k}_{1}$ and in-plane angle $\theta$. At the boundary, the electron may be retroreflected as a hole with momentum $\bm{k}_{1}$ or specularly reflected as an electron with momentum $\bm{k}_{2}$, corresponding to scattering events $A$ and $B$ in Fig.~\ref{schematic} respectively. In the basis of Eq. \eqref{bdgpsi} and setting $H(\bm{k}) = \hbar v_{\rm F}(\bm{\sigma}\times\bm{k})_{z} - \mu_{L}$, the wave function is given by
\begin{equation}
\begin{aligned}
\psi_{\rm I}(\bm{r}) &= \dfrac{e^{i\bm{k}_{1}^{+}\cdot \bm{r}}}{\sqrt{2}}\begin{pmatrix}
1 \\ -ie^{i\theta} \\ 0 \\ 0
\end{pmatrix}
 + A \dfrac{e^{i\bm{k}_{1}^{-}\cdot \bm{r}}}{\sqrt{2}}\begin{pmatrix}
 0 \\ 0 \\ 1 \\ -ie^{-i\theta}
 \end{pmatrix} 
 \\ & \ + B \dfrac{e^{i\bm{k}_{2}^{+}\cdot \bm{r}}}{\sqrt{2}}\begin{pmatrix}
1 \\ ie^{-i\theta} \\ 0 \\ 0
\end{pmatrix}.
\end{aligned}
\end{equation}
Note that we choose to use the spin basis of Eq.~\eqref{bdgpsi} since the helicity basis of Eq.~\eqref{bdgphi} is momentum dependent. Here $\bm{k}^{\pm} = (k_{x}^{+},k_{y})^{T}$, where $k_{x}^{\pm} = k_{x} \pm E/v_{Fx}(\bm{k})$. In region II the incident electron can either be transmitted as an electron-like quasiparticle with momentum $\bm{k}_{1}^{\prime}$ and in-plane angle $\theta^{\prime}$ or a hole-like quasiparticle with momentum $\bm{k}_{2}^{\prime} = (-k_{x}^{\prime},k_{y})^{T}$, corresponding to scattering events $C$ and $D$ in Fig.~\ref{schematic} respectively. Here we note that due to the translational invariance along the $y$-direction, the $k_{y}$ momentum is conserved.  The nonhelical wave function in region II is then
\begin{equation}
\begin{aligned}
\psi_{\rm II}(\bm{r}) &=C\dfrac{e^{i\bm{k}_{1}^{\prime +}\cdot \bm{r}}}{\sqrt{N_{-}}}\begin{pmatrix}
\cos(\vartheta_{\bm{c}(\bm{k}_{1}')}/2) \\ e^{i\varphi_{\bm{c}(\bm{k}_{1}')}}\sin(\vartheta_{\bm{c}(\bm{k}_{1}')}/2) \\ f_{-}(\bm{k}'_{1}) \sin(\vartheta_{\bm{c}(\bm{k}_{1}')}/2) \\ -f_{-}(\bm{k}'_{1}) e^{-i\varphi_{\bm{c}(\bm{k}_{1}')}}\cos(\vartheta_{\bm{c}(\bm{k}_{1}')}/2)
\end{pmatrix} 
\\ & \ + D \dfrac{ e^{i\bm{k}_{2}^{\prime -}\cdot \bm{r}}}{\sqrt{N_{+}}}\begin{pmatrix}
\cos(\vartheta_{\bm{c}(\bm{k}_{2}')}/2) \\ e^{i\varphi_{\bm{c}(\bm{k}_{2}')}}\sin(\vartheta_{\bm{c}(\bm{k}_{2}')}/2) \\ f_{+}(\bm{k}'_{2}) \sin(\vartheta_{\bm{c}(\bm{k}_{2}')}/2) \\ - f_{+}(\bm{k}'_{2}) e^{-i\varphi_{\bm{c}(\bm{k}_{2}')}}\cos(\vartheta_{\bm{c}(\bm{k}_{2}')}/2)
\end{pmatrix}.
\end{aligned}
\end{equation}
Here $N_{\pm}^{-1/2} = \Delta_{0}/\sqrt{2E(E \pm \sqrt{E^{2} - \Delta_{0}^{2})}}$ and $f_{\pm}(\bm{k}) = (E \pm \sqrt{E^{2} - \Delta_{0}^{2})}/(-e^{-i\varphi_{\bm{c}(\bm{k})}}\Delta_{0})$. Once again $\vartheta_{\bm{c}(\bm{k})}$ and $\varphi_{\bm{c}(\bm{k})}$ are the polar and azimuthal angles of the vector $\bm{c}(\bm{k})$ respectively. We have that $\bm{k}^{\prime \pm} = (k_{x}^{\prime \pm},k_{y})^{T}$ where $k_{x}^{\prime \pm} = k_{x}' \pm \sqrt{E^{2} - \Delta_{0}^{2}}/v_{Fx}(\bm{k}')$. As in the previous section the outgoing angle $\theta^{\prime}$ may be solved for in terms of the incoming angle $\theta$ due to the conservation of the $k_{y}$ momentum and the conservation of the normal state energy, noting that $\xi_{L}(\bm{k}) = \hbar v_{\rm F}|\bm{k}| - \mu_{L}$ in region I and $\xi_{R}(\bm{k}) = |\bm{c}(\bm{k})| - \mu_{R}$ in region II.

To solve the scattering problem we then must match the wave functions at the $x = 0$ boundary. Once again we can generally write
\begin{equation}
\psi_{\rm I}(0,y) = \mathcal{M}_{S}\psi_{\rm II}(0,y).
\label{Sboundarymatching}
\end{equation}
\begin{figure}
\includegraphics[width=\columnwidth]{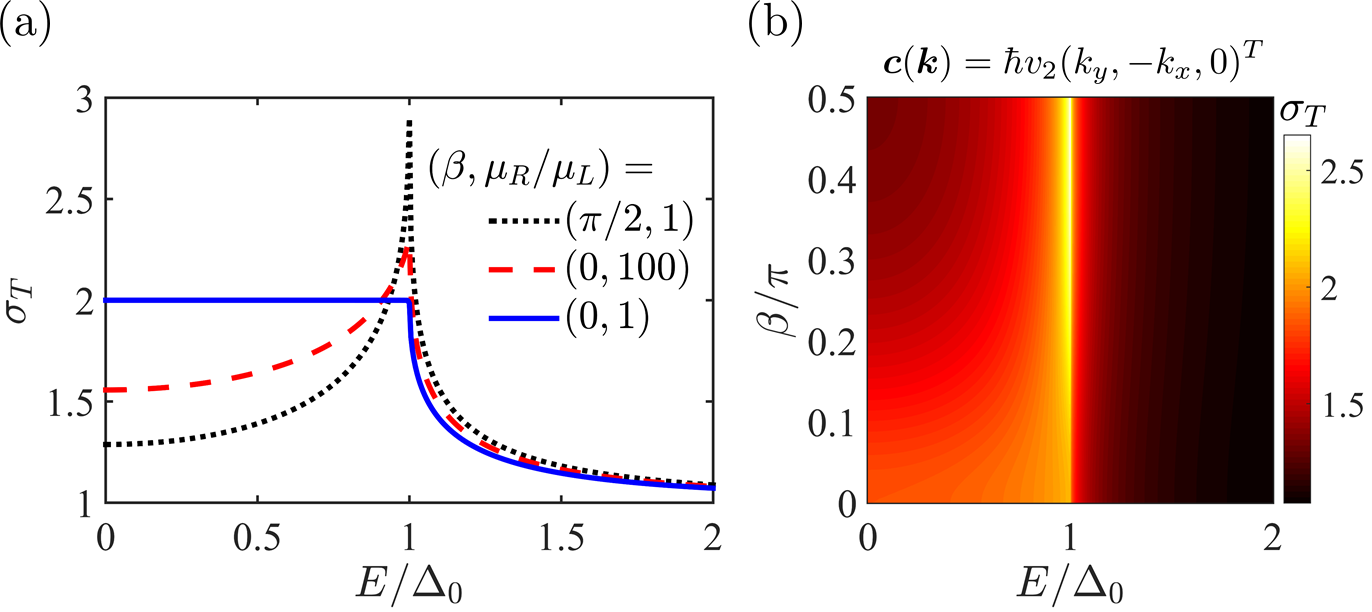}
\caption{(a) Total conductance $\sigma_{T}(E,\beta)$ as given by Eq.~\eqref{normtotcond} of the superconducting lateral heterojunction for a parent $s$-wave SC as $\bm{c}(\bm{k}) = \bm{c}_{\rm D}(\bm{k})$. Here it can be observed that a mismatch in the chemical potentials can introduce a peak at the gap edge, but not as large as the consequences of the angle-dependent scattering amplitude controlled by $\beta$ localized at the $x = 0$ boundary. (b) Total conductance $\sigma_{T}(E,\beta)$ of the superconducting lateral heterojunction for a parent $s$-wave SC in the case that the Fermi velocity in region II is reduced, such that $v_{2} = 0.7 v_{\rm F}$.}
\label{chem}
\end{figure}
In the mean field BdG description the Hamiltonian $\mathcal{H}$ is written in a redundant formalism where the hole degrees of freedom are constructed from the electron Hamiltonian $H(\bm{k})$ by writing $-H^{T}(-\bm{k})$. Following the methods of Sec.~\ref{sec2} we find that the boundary value matrix for the hole Hamiltonian $-H^{T}(-\bm{k})$ is given by $\mathcal{M}^{*}(\beta)$. Because $-H^{T}(-\bm{k})$ is not independent of $H(\bm{k})$, we see that if $\mathcal{M}(\beta)$ has a particular value for its free parameter $\beta$, then $\mathcal{M}^{*}(\beta)$ must also have this same value for its free parameter. The boundary value matrix for the BdG wave functions is then
\begin{equation}
\mathcal{M}_{S}(\beta) = \begin{pmatrix}
\mathcal{M}(\beta) & 0
\\ 0 & \mathcal{M}^{*}(\beta)
\end{pmatrix}.
\end{equation}
The block diagonal nature of the matrix $\mathcal{M}_{S}(\beta)$ means that there is no additional particle hole mixing due to the boundary conditions, which is a natural assumption.

Matching the wave functions by writing $\psi_{\rm I}(0,y) = \mathcal{M}_{S}(\beta)\psi_{\rm II}(0,y)$ and solving for the coefficients $A$ and $B$, we then define the transmission coefficient of the superconducting lateral heterojunction~\cite{Blonder1982,Kashiwaya1996,Tanaka1995} as $\sigma_{S}(E,\theta,\beta) = 1 + |A|^{2} - |B|^{2}$. From here, we can define the normalized conductance and the total conductance respectively as
\begin{equation}
\begin{aligned}
&\sigma_{R}(E,\theta,\beta) = \dfrac{\sigma_{S}(E,\theta,\beta)}{\sigma_{N}(\theta,\beta)},
\\ &\sigma_{T}(E,\beta) = \dfrac{\int_{-\pi/2}^{\pi/2}\sigma_{S}(E,\theta,\beta)\cos\theta d\theta}{\int_{-\pi/2}^{\pi/2}\sigma_{N}(\theta,\beta)\cos\theta d\theta},
\end{aligned}
\label{normtotcond}
\end{equation}
where $\sigma_{N}(\theta,\beta)$ is the normal state conductance given in Eq. \eqref{sigman}.

\begin{figure}
\includegraphics[width=\columnwidth]{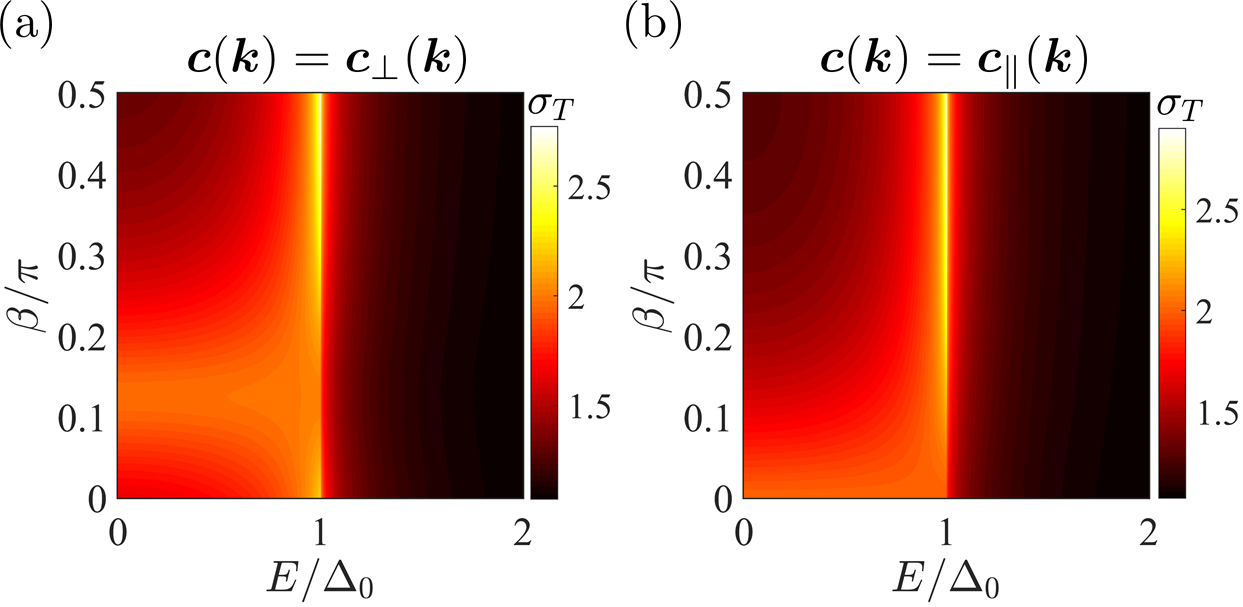}
\caption{(a) Total conductance $\sigma_{T}(E,\beta)$ as given by Eq.~\eqref{normtotcond} of the superconducting lateral heterojunction in the case that $\bm{c}(\bm{k}) = \bm{c}_{\perp}(\bm{k})$ in region II, where $\bm{c}_{\perp}(\bm{k}) = \hbar v_{\rm F}(\lambda k_{y},-k_{x},-\lambda k_{y})^{T}$. (b) Total conductance in the case that $\bm{c}(\bm{k}) = \bm{c}_{\parallel}(\bm{k})$ in region II, where $\bm{c}_{\parallel}(\bm{k}) = \hbar v_{\rm F}(k_{y},-\lambda k_{x},-\lambda k_{x})^{T}$. Here both $\bm{c}_{\perp}(\bm{k})$ and $\bm{c}_{\parallel}(\bm{k})$ are defined in Eq.~\eqref{cperppara}, and for both cases we have a parent $s$-wave SC and $\lambda = 2/3$.}
\label{elliptical}
\end{figure}

\section{Conductance of the Lateral Heterojunction}
\label{sec4}
\begin{figure*}
\includegraphics[width=\textwidth]{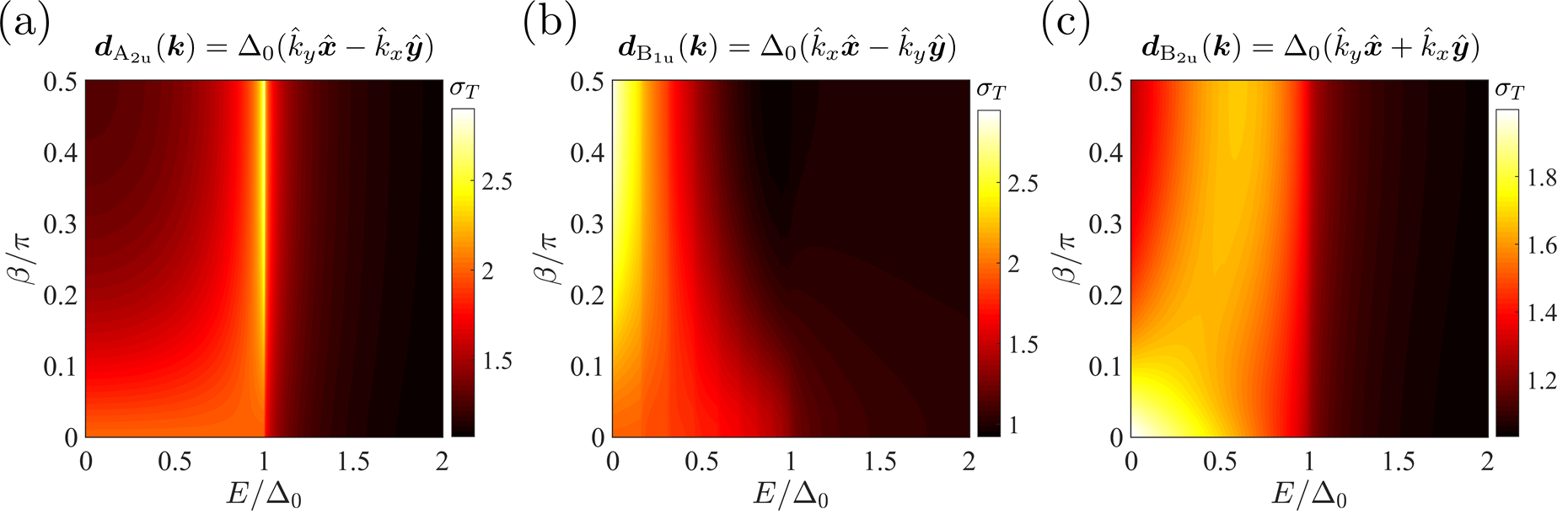}
\caption{Total conductance $\sigma_{T}(E,\beta)$ as given by Eq.~\eqref{normtotcond} for parent helical spin-triplet SCs according to the $D_{4\rm h}$ point group in the case that $\bm{c}(\bm{k}) = \bm{c}_{\rm D}(\bm{k})$ for (a) a parent $\rm A_{2\rm u}$ SC (b) a parent $\rm B_{1\rm u}$ SC (c) a parent $\rm B_{2\rm u}$ SC.}
\label{triplets}
\end{figure*}
In this section we analyze the consequences of nonhelical spin textures, as encoded in the choice of $c_{ij}$ coefficients, on the conductance of superconducting lateral heterostructures with parent $s$-wave SCs. We first analyze the normalized conductance $\sigma_{R}$ in the case that $\bm{c}(\bm{k}) = \bm{c}_{\rm D}(\bm{k})$ in region II, where $\bm{c}_{\rm D}(\bm{k}) = \hbar v_{\rm F}(k_{y},-k_{x},0)^{T}$ so that both sides of the heterojunction admit helical surface states. We find that when $\beta = 0$ there is no scattering at the junction and all incoming angles experience perfect Andreev reflection for energies below the gap edge. When $\beta = \pi/2$ the probability of specular reflection due to scattering increases, resulting in a suppressed conductance for energies below the gap edge as shown in Fig.~\ref{normcond}(a). In the standard treatment by Blonder \textit{et. al.}, as the scattering at the junction increases the conductance evolves from the Andreev limit to the profile reflecting the tunneling density of states.~\cite{Blonder1982} Note that this is not what takes place with increasing $\beta$: the low energy conductance never vanishes, as seen in the total conductance in Fig.~\ref{normcond}(b) and the black dotted line in Fig.~\ref{chem}(a). This is because the normal incidence angle always experience perfect Andreev reflection regardless of the strength of $\beta$, reflecting the inability to confine Dirac particles due to Klein tunneling. As in the normal state, Fig.~\ref{normcond}(a) shows that $\beta$ acts as an angle-dependent scattering amplitude. In Fig.~\ref{normcond}(b) we plot the total conductance for all $\beta\in[0,\pi/2)$. For $\beta = 0$ we see that we have a constant $\sigma_{T}(E,0) = 2$ for all $E<\Delta_{0}$, followed by a decreasing conductance for larger energies above the superconducting gap. Whenever $\beta \neq 0$ the conductance instead finds a maximum value at the $E = \Delta_{0}$ gap edge. For conventional $s$-wave SCs in a normal-superconducting junction, localized interface potentials at the boundary can result in divergences of the total conductance at the gap edge.\cite{Blonder1982} This is in contrast to the Dirac Hamiltonians that describe our lateral heterojunction, as the total conductance only obtains a maximum which is largest at the gap edge for $\beta = \pi/2$. The concomitant result is that the total conductance has a maximum, rather than a singularity, at the gap edge.

In Fig.~\ref{chem}(a), we plot the total conductance for the case $\bm{c}(\bm{k}) = \bm{c}_{\rm D}(\bm{k})$ and compare the effects of the angle-dependent scattering controlled by $\beta$ to that of a mismatch of the chemical potentials between regions I and II. While $\mu_{L} \neq \mu_{R}$ similarly creates a maximum at the gap edge, the effect is not as pronounced as that arising from setting $\beta = \pi/2$. In Fig.~\ref{chem}(b) we plot the total conductance in the case that region II has a renormalized Fermi velocity such that $\bm{c}(\bm{k}) = \hbar v_{2}(k_{y},-k_{x},0)$, where $v_{2} = 0.7v_{\rm F}$. The reduced Fermi velocity creates a mismatch in the size of the Fermi surfaces between regions I and II, leading to extra scattering in addition to that controlled by $\beta$.


Nonhelical spin textures in region II that result from rotational symmetry breaking interface potentials within TI-SC heterostructures can introduce elliptical energy contours and out-of-plane spin textures within the TI interface states.\cite{Alspaugh2018,Asmar2017} Two generic examples of such nonhelical spin textures can be modeled by 
\begin{equation}
\begin{aligned}
\bm{c}_{\perp}(\bm{k}) &= \hbar v_{\rm F}(\lambda k_{y},-k_{x},-\lambda k_{y})^{T},
\\ \bm{c}_{\parallel}(\bm{k}) &= \hbar v_{\rm F}( k_{y},-\lambda k_{x},-\lambda k_{x})^{T},
\end{aligned}
\label{cperppara}
\end{equation}
where $0 < \lambda < 1$.~\cite{Alspaugh2018} The former spin texture introduces an elliptical Fermi surface with a major axis perpendicular to the $x$-axis, while the latter spin texture has an elliptical Fermi surface with a major axis parallel to the $x$-axis. In Figs.~\ref{elliptical}(a) and \ref{elliptical}(b) we plot the total conductance for each of these two cases respectively. Comparing Fig.~\ref{normcond}(b) and Fig.~\ref{elliptical}(b), we see that $\bm{c}_{\parallel}(\bm{k})$ has the same total conductance spectrum as the helical surface states of $\bm{c}_{\rm D}(\bm{k})$, due to the fact that the conservation of the $k_{y}$ momentum is unaffected by the ellipticity in this direction. In contrast, the total conductance of $\bm{c}_{\perp}(\bm{k})$ is modified compared to that of the helical surface states. However, for $\beta = \pi/8$ in Fig.~\ref{elliptical}(a) the scattering of the $x = 0$ boundary is reduced and the system experiences perfect Andreev reflection. It follows that the direction of the ellipticity of the Fermi surface cannot be distinguished from the consequences of $\beta$ as different choices of $\beta$ can produce similar conductance spectra between $\bm{c}_{\perp}(\bm{k})$ and $\bm{c}_{\parallel}(\bm{k})$. In addition, we find that different ellipse directions require different values of $\beta$ to maximize the normal state conductance.


\section{Spin-Triplet Superconductors}
\label{sec5}
Next we study the conductance signatures for several examples of parent spin-triplet SCs. To analyze these systems, we replace the order parameter of the parent SC in Eq.~\eqref{bdgpsi} with $\hat{\Delta} = [\bm{d}(\bm{k})\cdot\bm{\sigma}](i\sigma_{y})$, where the vector $\bm{d}(\bm{k})$ defines the spin-triplet part of the pairing field. Carrying through a similar analysis as before, it has been shown in Ref.~\onlinecite{Alspaugh2018} that the proximity-induced order parameter of the TI-SC interface in Eq.~\eqref{bdgphi} now takes the form $\Delta(\bm{k}) = -e^{-i\varphi_{\bm{c}(\bm{k})}}[\hat{\bm{c}}(\bm{k})\cdot\bm{d}(\bm{k})]$, where $\hat{\bm{c}}(\bm{k}) = \bm{c}(\bm{k})/|\bm{c}(\bm{k})|$. 

For concreteness we assume a tetragonal crystal symmetry for the parent superconducting material and classify $\bm{d}(\bm{k})$ according to the irreducible representations of the $D_{4\rm h}$ point group. The conductance signatures of the chiral $\rm E^{\pm}_{2\rm u}$ pairing states in which $\bm{d}_{\rm E^{\pm}_{2\rm u}}(\bm{k}) = \Delta_{0}(\hat{k}_{x} \pm i\hat{k}_{y})\hat{\bm{z}}$ have been studied in Ref.~\onlinecite{Alspaugh2018}, and below we focus on the helical $\rm A_{1\rm u}$, $\rm A_{2\rm u}$, $\rm B_{1\rm u}$, and $\rm B_{2\rm u}$ pairing states.

Assuming that $\bm{c}(\bm{k}) = \bm{c}_{\rm D}(\bm{k})$ in region II and thus $\Delta_{\rm D}(\bm{k}) = -ie^{-i\theta}[\bm{d}(\bm{k})\times\hat{\bm{k}}]_{z}$, we find that the $\rm A_{2\rm u}$ pairing state with $\bm{d}_{\rm A_{2\rm u}}(\bm{k}) = \Delta_{0}(\hat{k}_{y}\hat{\bm{x}} - \hat{k}_{x}\hat{\bm{y}})$ produces isotropic fully gapped superconductivity at the interface. In contrast, the $\rm B_{1\rm u}$ pairing state with $\bm{d}_{\rm B_{1\rm u}}(\bm{k}) = \Delta_{0}(\hat{k}_{x}\hat{\bm{x}} - \hat{k}_{y}\hat{\bm{y}})$ and the $\rm B_{2\rm u}$ pairing state with $\bm{d}_{\rm B_{2\rm u}}(\bm{k}) = \Delta_{0}(\hat{k}_{y}\hat{\bm{x}} + \hat{k}_{x}\hat{\bm{y}})$ both produce $d$-wave-like nodal gaps. These differences manifest themselves in the Andreev conductance spectra plotted in Fig.~\ref{triplets}. For the $\rm B_{2\rm u}$ pairing state at $\beta = 0 $ there is a bright spot corresponding to Andreev reflection near $E = 0$. That peak is shifted towards the finite in-gap value as $\beta$ increases. In contrast, for the $\rm B_{1\rm u}$ pairing state there is a zero energy peak in the total conductance for all values of $\beta$.

\begin{figure}
\includegraphics[width=\columnwidth]{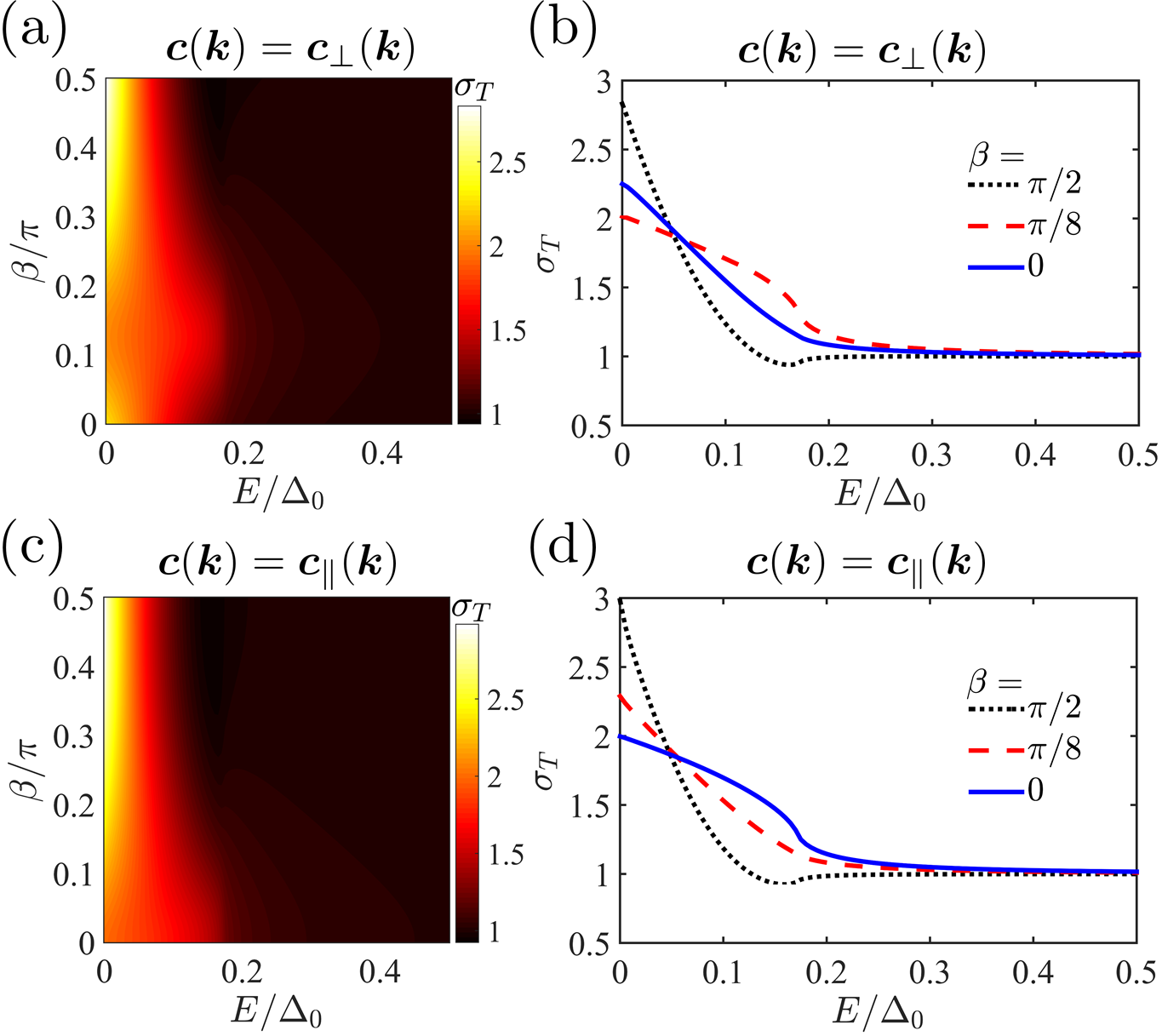}
\caption{Total conductance $\sigma_{T}(E,\beta)$ as given by Eq.~\eqref{normtotcond} for an $\rm A_{1\rm u}$ parent helical spin-triplet SC as defined by the $D_{4\rm h}$ point group. In (a) and (c) we plot the total conductance for the nonhelical spin textures described by $\bm{c}_{\perp}(\bm{k})$ and $\bm{c}_{\parallel}(\bm{k})$ as given in Eq.~\eqref{cperppara} respectively. In (b) and (d) we plot the same conductance spectra for chosen values of $\beta$.}
\label{A1u}
\end{figure}
For the $\rm A_{1\rm u}$ pairing state with $\bm{d}_{\rm A_{1\rm u}}(\bm{k}) = \Delta_{0}(\hat{k}_{x}\hat{\bm{x}} + \hat{k}_{y}\hat{\bm{y}})$ we find that the order parameter vanishes and there is no proximity-induced superconductivity when $\bm{c}(\bm{k}) = \bm{c}_{\rm D}(\bm{k})$. For this family of spin-triplet SCs the proximity effect thus requires the presence of nonhelical spin textures in the TI-SC barrier. In Fig.~\ref{A1u} we plot the total conductance for the nonhelical spin textures described by $\bm{c}_{\perp}(\bm{k})$ and $\bm{c}_{\parallel}(\bm{k})$ respectively. In both cases we find that the conductance always obtains its largest value at zero energy regardless of the value of $\beta$. As before, the orientation of the elliptical energy contours cannot be distinguished from the effects of scattering as different values of $\beta$ give rise to similar conductance spectra. We calculate that at energies comparable to the proximity-induced gap edge, the conductance tends towards unity. While we cannot test for the direction of the dispersion's ellipticity, the existence of the anisotropy would be proven by the observation of the signal.

\section{Discussion}
\label{sec6}
In this work we analyzed how nonhelical spin textures present in TI-SC interface states affect the conductance of lateral heterojunctions, which can be created by placing a parent SC in contact with only half of a TI surface as shown schematically in Fig.~\ref{schematic}. It has been shown that TI-vacuum terminations host helical surface states such that the total angular momentum of the propagating state is perpendicular to its velocity and confined within the plane of the surface.~\cite{Hasan2010,Qi2011} In contrast, TIs in contact with nontopological materials host interface states that can exhibit elliptical energy contours and nonhelical spin textures that point out of the plane of the interface.~\cite{Alspaugh2018,Asmar2017}

We evaluated the conductance of the system in the normal state and found that angle-dependent scattering amplitudes localized at the one-dimensional boundary between the TI-vacuum surface (region I) and the TI-SC interface (region II), as shown schematically in Fig.~\ref{schematic}, lead to a reduction of the current through the heterojunction. This reduction is largely due to the scattering of near-grazing angles at the boundary between regions I and II. Introducing $s$-wave superconductivity, we then found that both the angle-dependent scattering amplitude and a mismatch of the chemical potentials between regions I and II can both reduce the Andreev conductance through the device, with the former leading to a more dramatic suppression. We also calculated the total conductance for a variety of nonhelical spin textures, which all lead to qualitatively similar conductance spectra for parent $s$-wave SCs characterized by a finite, albeit reduced, conductance at $E = 0$ and a peak at the gap edge. Finally, we then computed the total conductance spectra for parent spin-triplet SCs in region II as classified according to the irreducible representations of the $D_{4\rm h}$ point group. In particular, we find that for parent helical $\rm A_{1\rm u}$ SCs, nonzero conductance spectra only exist when nonhelical spin textures are present within region II.

Our results in general demonstrate that nonhelical spin textures have important and observable consequences for TI-based devices. TI-SC heterostructures placed alongside ferromagnetic materials have earlier been proposed to host Majorana zero modes as a consequence of the spin-momentum locking intrinsic to topological interface states. In this article we show that the spin-momentum locking within heterojunctions is more complex than has been commonly assumed, and leads to a more diverse set of phenomena even for non-magnetic heterostructures. Our work therefore constitutes a crucial step towards understanding the role of  non-helical spin states in such heterostructures.


\section{Acknowledgments}
This research was supported by NSF via Grants No. DMR-1410741 (D.J.A., M.M.A., and I.V.) and No. DMR-1151717 (M.M.A. and D.E.S.).

\bibliography{bibfile} 

\bibliographystyle{apsrev4-1}

\end{document}